\documentclass[lettersize,journal]{IEEEtran}
\usepackage{amsmath,amsfonts}
\usepackage{algorithmic}
\usepackage{algorithm}
\usepackage{array}
\usepackage[caption=false,font=normalsize,labelfont={sf,scriptsize},textfont=sf]{subfig}
\usepackage{textcomp}
\usepackage{stfloats}
\usepackage{url}
\usepackage{verbatim}
\usepackage{graphicx}
\usepackage{cite}
\usepackage{amsmath,amssymb,amsfonts}
\usepackage{arydshln}
\newtheorem{definition}{\it Definition}
\newtheorem{assumption}{\it Assumption}
\newtheorem{theorem}{\it Theorem}
\newtheorem{lemma}{\it Lemma}

\begin{document}

\title{Similar Formation Control of Multi-Agent Systems over Directed Acyclic Graphs via Matrix-Weighted Laplacian}
\author{Zhipeng Fan,
Yujie Xu,
Mingyu Fu,
Han Sun,
Weiqiu Zhang,
Heng Zhang
\thanks{This work was produced by the National Natural Science Foundation of China under Grant 52071112.(Corresponding author: Yujie Xu.)}
\thanks{Zhipeng Fan, Yujie Xu, Mingyu Fu, Han Sun and Weiqiu Zhang are with the College of Intelligent Systems Science and Engineering, Harbin Engineering University, Harbin 150001,
 China (e-mail: chrisfan@hrbeu.edu.cn; xuyujie@hrbeu.edu.cn; fumingyu@hrbeu.edu.cn; heuzhangweiqiu@163.com; sunhan65@hrbeu.edu.cn
). }
\thanks{Heng Zhang is with the Sanya Nanhai Innovation and Development Base, Harbin Engineering University, Sanya, 572000, China (e-mail: zhh@hrbeu.edu.cn 
). }
}

\markboth{Journal of \LaTeX\ Class Files,~Vol.~14, No.~8, August~2021}%
{Shell \MakeLowercase{\textit{et al.}}: A Sample Article Using IEEEtran.cls for IEEE Journals}

\IEEEpubid{}

\maketitle

\begin{abstract}
This brief proposes a distributed formation control strategy via matrix-weighted Laplacian that can achieve a similar formation in 2-D planar using inter-agent relative displacement measurement. Formation patterns that include translation, rotation, and scaling can be characterized by the null space of the matrix-weighted Laplacian associated with the topological graph. The main contribution of this brief is to extend the similar formation problem of undirected graphs to directed acyclic graphs and provide the necessary algebraic criteria for leader selection. Stability analysis, illustrative examples, and simulation results are provided.
\end{abstract}


\section{Introduction}
\label{sec:introduction}

\IEEEPARstart{T}{he} fascinating collective behaviors of biological systems have inspired extensive studies on group control of multi-agent\cite{zhao2015bearing}, \cite{sun2023mean},\cite{li2023prescribed}. Formation control is an important direction in the field of group control and has many applications in several areas such as search and rescue missions, environmental mapping and monitoring \cite{sun2023mean}, \cite{gunn2015dynamic}, \cite{fan2025two}.

Depending on the form of the Laplacian matrix, existing works can be mainly classified into standard Laplacian matrix \cite{zhang2023distributed},\cite{fan2024unified}, bearing Laplacian matrix \cite{li2023prescribed}, \cite{zhao2016localizability}, \cite{li2022bearing}, complex Laplacian matrix \cite{lin2013leader},\cite{lin2014distributed}, \cite{fang2023distributed}, \cite{fang2023distributed2} and signed Laplacian matrix \cite{lin2015necessary}, \cite{zhao2018affine}, \cite{zhi2021leader}. The standard Laplacian can only perform translational maneuvers because of the limited null space of its Laplacian matrix, and the bearing Laplacian can achieve translational and scaling maneuvers. Conversely, the signed Laplacian matrix possesses the highest dimension of null space, thereby facilitating versatile transformations such as translation, scaling, rotation, and shear (i.e., affine transformations) of the configuration. Despite the signed Laplacian matrix having an advantage in formation shape transformation, each follower is required to have at least three neighbors, which means that the topology network will be very complicated. As a trade-off between different Laplacian matrices, the complex Laplacian has a relatively simple topology network while also being able to flexibly implement translation, rotation, and scaling (i.e., similar transformations) of the formation.
Due to the above virtues of complex Laplacian, it has been a popular topic recently \cite{fang2023distributed}, \cite{fang2023distributed2}. It is worth noting that in complex space, the Lyapunov stability analysis theory cannot be directly applied \cite{fang2023distributed}, which limits the generalisation of the complex Laplace formation theory in the field of nonlinear systems. Fortunately, the complex Laplacian-based approach has been generalized to the real space by reconstructing the complex weight as a matrix-weighted value \cite{fathian2017distributed}. Existing results have explored the matrix-weighted Laplacian formation control over undirected graphs \cite{fathian2017distributed}, \cite{fathian2020robust}, \cite{xu2021distributed}, but, for directed graphs, it is still an unsolved problem to a large extent up to now. On the basis of the observation, a basic framework for a similar formation issue in directed acyclic graphs will be established by our further work.

In this brief, we explore the theory of similar formation for directed acyclic graphs (a class of minimally acyclic persistent graphs) based on matrix-weighted Laplacian, which is an extension of existing results\cite{fathian2017distributed}, \cite{fathian2020robust}, \cite{xu2021distributed}. The main contributions of this brief are threefold. First, we provide the algebraic criteria for leader selection, which determines whether the selected leaders can fully control the entire formation to realize desired similar transformations. Second, based on the idea of topology renumbering, considering a class of minimally acyclic persistent graphs, we derive both algebraic and graphical conditions for similar localizability. That is, similar formation problems over undirected graphs are extended to directed acyclic graphs in this brief. Third, we address similar formation issue of multi-agent systems with single-integrator dynamics using the linear system stability theory.

\section{Preliminaries}
\subsection{Notations for formations}
Throughout the article, let $\mathbb{R}$ be the set of real. Let $\operatorname{Null}(\cdot)$ and Range$(\cdot)$ be the null and column spaces of a matrix, respectively. Let $\|\cdot\|$ be the Euclidian norm of a vector or the spectral norm of a matrix.

Consider a team of $n$ nodes in real space $\mathbb{R}^2$. Denote the first $n_l$ nodes as leaders and the rest $n_f=n-n_l$ nodes as followers. Thus, we define the set of all nodes as $\mathcal{V}=\{1, \ldots, n\}$, the set of leader nodes as $\mathcal{V}_l=\left\{1, \ldots, n_l\right\}$, and the set of follower nodes as $\mathcal{V}_f=\mathcal{V} \backslash \mathcal{V}_l$. We use the graph $\mathcal{G}=(\mathcal{V}, \mathcal{E})$ to characterize the network topology among the nodes, where $\mathcal{E} \subseteq \mathcal{V} \times \mathcal{V}$ is the set of edges. $(i, j) \in \mathcal{E}$ indicates that node $i$ can receive information from node $j$ and $\mathcal{N}_i=\{j \in \mathcal{V} \mid(i, j) \in \mathcal{E}\}$ denotes the set of neighbors of node $i$. A formation $\mathcal{G}(\mathcal{V}, p)$ is the graph $\mathcal{G}$ with its node $i$ mapped to $p_i$. The leaders' positions are $p_l=\left[p_1^T, \ldots, p_{n_l}^T\right]^T$ and the followers' positions are $p_f=\left[p_{n_{l+1}}^T, \ldots, p_n^T\right]^T$, respectively, so $p=\left[p_l^T, p_f^T\right]^T$. 

\subsection{Equilibrium condition}
First, we introduce the following matrix-weighted value attributed on edge $(i, j)$ \cite{fathian2020robust}
\begin{equation}
\label{Eq:matrix-weighted value}
\mathcal{B}_{i j}=\left[\begin{array}{@{}cc@{}}
a_{i j} & -b_{i j} \\
b_{i j} & a_{i j}
\end{array}\right]=c_{i j} R_{i j} 
\end{equation}
 where $ a_{i j}, b_{i j} \in \mathbb{R} $ and $c_{i j}=\sqrt{a_{i j}^2+b_{i j}^2} \neq 0. R_{i j}=c_{i j}^{-1} \mathcal{B}_{i j} \in S O(2) $ is the 2-D rotation matrix.

\begin{figure}[!t]
\centering
\includegraphics[width=1.8in]{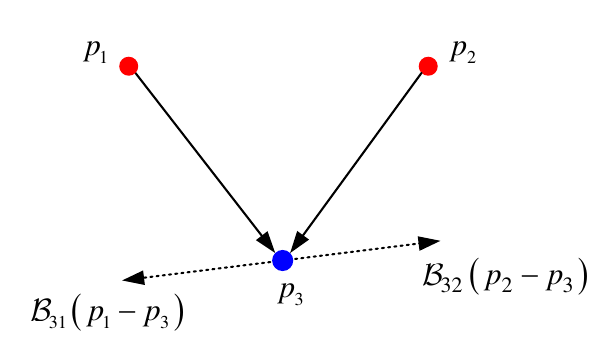}
\caption{Illustration of equilibrium condition.}
\label{Fig:Illustration of equilibrium condition}
\end{figure}

For formation $(\mathcal{G}, p)$, a linear constraint is called equilibrium condition if it satisfies
\small {
\begin{equation}
\label{Eq:equilibrium condition}
\begin{aligned}
&\sum_{j \in \mathcal{N}_i} \mathcal{B}_{i j}\left(p_i-p_j\right) \\
=&\sum_{j \in \mathcal{N}_i} c_{i j} R_{i j}\left(p_i-p_j\right)=0, \quad i \in \mathcal{V}_f  
\end{aligned}
\end{equation}   }

The $c_{i j}$ and $R_{i j}$ respectively make the relative positions vectors scaled and rotated so that the equilibrium condition (\ref{Eq:equilibrium condition}) holds (see Fig. \ref{Fig:Illustration of equilibrium condition} for an illustration).

As demonstrated in Fig. \ref{Fig:Illustration of equilibrium condition}, the equilibrium condition (\ref{Eq:equilibrium condition}) can be achieved with only two neighbors in the network. Based on the observation, we give the following assumption.
\begin{assumption}
\label{ass:neighbors number}
Each follower has only two neighbors.
\end{assumption}

Then, a matrix-weighted Laplacian for graph $\mathcal{G}$ is defined as
\small {
\begin{equation}
\label{Eq:matrix-weighted Laplacian}
\mathcal{L}(i, j)= \begin{cases}-\mathcal{B}_{i j} & \text { if } i \neq j \text { and } j \in \mathcal{N}_i \\ \mathbf{0}_{2 \times 2} & \text { if } i \neq j \text { and } j \notin \mathcal{N}_i \\ \sum_{k \in \mathcal{N}_i} \mathcal{B}_{i k} & \text { if } i=j\end{cases}
\end{equation}   }

We can aggregate the linear constraint from (\ref{Eq:equilibrium condition}) into a compact form as 
\begin{equation}
\label{Eq:linear constraint a compact form}
\mathcal{L} \cdot p=0
\end{equation}
where $\mathcal{L}$ is referred to as matrix-weighted Laplacian defined in (\ref{Eq:matrix-weighted Laplacian}).

\begin{figure}[!t]
\centering
\includegraphics[width=2.2in]{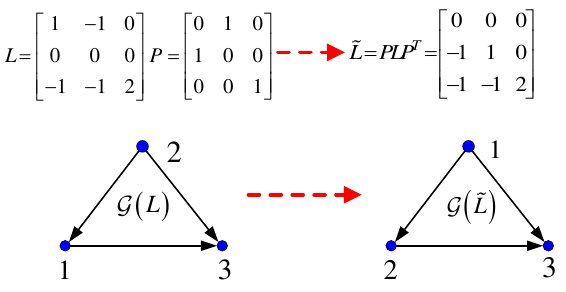}
\caption{Renumbering the nodes and obtaining lower triangular Laplacian matrix.}
\label{Fig:Renumbering the nodes}
\end{figure}

\begin{lemma}[\cite{cao2018group}]
\label{lem:reorder topology}
For any directed acyclic graph $\mathcal{G}$, by a permutation similarity transformation (i.e., renumbering the nodes if necessary), the standard Laplacian matrix $L$ corresponding to $\mathcal{G}$ can be rewritten as the following lower triangular form
\small{
\begin{equation*}
\tilde{L}=P L P^T=\left[\begin{array}{@{}cccc@{}}
\tilde{l}_{11} & 0 & \ldots & 0 \\
\tilde{l}_{21} & \tilde{l}_{22} & \ldots & 0 \\
\vdots & \vdots & \ddots & \vdots \\
\tilde{l}_{n 1} & \tilde{l}_{n 2} & \ldots & \tilde{l}_{n n}
\end{array}\right]
\end{equation*}
}
where $P$ is a permutation matrix (see \ref{Fig:Renumbering the nodes} for an illustration).
\end{lemma}

\section{Problem statement and design of similar framework}
\subsection{Target formation and control objective}
Denote the time-varying configuration of the target formation as 
\begin{equation}
\label{Eq:the target formation}
p^*(t)=\left[I_n \otimes \alpha(t) \mathcal{R}(t)\right] r+\mathbf{1}_n \otimes b(t)
\end{equation}
where $\alpha(t) \in \mathbb{R^+}$ is the scaling factor, $\mathcal{R}(t)=\left[\begin{array}{@{}cc@{}}\cos (\vartheta(t)) & -\sin (\vartheta(t)) \\ \sin (\vartheta(t)) & \cos (\vartheta(t))\end{array}\right] \in S O(2)$ is a rotation matrix with $\vartheta(t) \in[0,2 \pi)$ and $b(t) \in \mathbb{R}^2 . r=\left[r_1^T, \ldots, r_n^T\right]^T=\left[r_l^T, r_f^T\right]^T \in \mathbb{R}^{2 n}$ is a constant nominal configuration to be designed later.

In this brief, we assume that all leaders do not interact with followers and do not need to access the information from followers. Hence, the Laplacian of graph $\mathcal{G}$ can take the following form \cite{lin2013leader}

\begin{equation}
\label{Eq:graph Laplacian}
\mathcal{L} = 
\left[\begin{array}{@{}c:c@{}}
\mathbf{0}_{2n_l \times 2n_l} & \mathbf{0}_{2n_l \times 2n_f} \\
 \hdashline \mathcal{L}_{f l} & \mathcal{L}_{f f}
\end{array}\right]
\end{equation}

The control objective is to steer all followers converge to their desired positions, and the target formation can be realized with maneuvers, i.e.,
\begin{equation}
\label{Eq:control objective 1}
\lim _{t \rightarrow \infty}\left(p_i(t)-p_i^*(t)\right)=0, \quad i \in \mathcal{V}_f
\end{equation}

According to Theorem \ref{theorem:Algebraic condition for similar localizability}, under the similar formation scheme, the desired positions of the follower can be uniquely calculated as $p_f^*=-\mathcal{L}_{f f}^{-1} \mathcal{L}_{f l} p_l$. Thus, the control objective (\ref{Eq:control objective 1}) can be rewritten as
\begin{equation}
\label{Eq:control objective 2}
\begin{aligned}
& \lim _{t \rightarrow \infty}\left(p_f(t)-p_f^*(t)\right) \\
= & \lim _{t \rightarrow \infty}\left(p_f(t)+\mathcal{L}_{f f}^{-1} \mathcal{L}_{f l} p_l(t)\right) \rightarrow 0
\end{aligned}
\end{equation}

\subsection{Similar image of nominal configuration}
The similar image of the nominal configuration is defined as \cite{xu2021distributed}
\begin{equation}
\label{Eq:similar image of the nominal configuration}
\begin{aligned}
\mathcal{S}(r)=\left\{p \in \mathbb{R}^{2n}: p=\right. & \left(I_n \otimes \alpha \mathcal{R}\right) r+\mathbf{1}_n \otimes b, \\
& \left.\alpha \in \mathbb{R^+}, \mathcal{R} \in S O(2), b \in \mathbb{R}^2\right\}
\end{aligned}
\end{equation}

The similar image is a set consisting of all the translation, rotation, and scaling transformations of the nominal configuration. Notice that the coefficient $b$ is related to translation, while $\alpha$ and $\mathcal{R}$ are related to scaling and rotation, respectively.

Next, we present a fundamental result regarding the similar image $\mathcal{S}(r)$.
\begin{lemma}[Dimension of similar image]
\label{lem:Dimension of similar image}
If $r_i \neq r_j$ for $i, j \in \mathcal{V}$ and $n \geq 2$ are satisfied, then the dimension of $\mathcal{S}(r)$ equals 4 and $\mathcal{S}(r)=\operatorname{span}\left\{r, r^{\prime}, \mathbf{1}_n \otimes I_2\right\}$, where $r^{\prime}$ is shown in (\ref{Eq:four dependent vectors}). 
\end{lemma}
\begin{IEEEproof}[Proof]
Denote 
\small {
\begin{equation}
E_1=\left[\begin{array}{@{}ll@{}}
1 & 0 \\
0 & 1
\end{array}\right], E_2=\left[\begin{array}{@{}cc@{}}
0 & -1 \\
1 & 0
\end{array}\right], e_1=\left[\begin{array}{@{}l@{}}
1 \\
0
\end{array}\right], e_2=\left[\begin{array}{@{}l@{}}
0 \\
1
\end{array}\right]
\nonumber
\end{equation}   }

Consider the following four vectors
\small{
\begin{equation}
\begin{aligned}
& r=\left(I_n \otimes E_1\right) r=\left[r_{11}, r_{12}, \ldots, r_{n 1}, r_{n 2}\right]^T \\
& r^{\prime}=\left(I_n \otimes E_2\right) r=\left[-r_{12}, r_{11}, \ldots,-r_{n 2}, r_{n 1}\right]^T \\
& \mathbf{1}^{\prime}=\mathbf{1}_n \otimes e_1=[1,0, \ldots, 1,0]^T \\
& \mathbf{1}^{\prime \prime}=\mathbf{1}_n \otimes e_2=[0,1, \ldots, 0,1]^T
\end{aligned}
\label{Eq:four dependent vectors}
\end{equation}
}

It is easy to verify that these vectors are all in $\mathcal{S}(r)$, and any other vectors in $\mathcal{S}(r)$ can be expressed as a linear combination of them by noting that the following fact
\small{
\begin{equation}
\begin{aligned}
& \left(I_n \otimes \alpha \mathcal{R}\right) r+\mathbf{1}_n \otimes b \\
= & \alpha\left(I_n \otimes \mathcal{R}_{11} E_1\right) r+\alpha\left(I_n \otimes \mathcal{R}_{21} E_2\right) r+b_1 \mathbf{1}^{\prime}+b_2 \mathbf{1}^{\prime \prime} \\
= & \alpha \mathcal{R}_{11} r+\alpha \mathcal{R}_{21} r^{\prime}+b_1 \mathbf{1}^{\prime}+b_2 \mathbf{1}^{\prime \prime}
\end{aligned}
\nonumber
\end{equation} }
where $\mathcal{R}_{11}$ and $\mathcal{R}_{21}$ represent $\mathcal{R}(1,1)$ and $\mathcal{R}(2,1)$, respectively. $b_1$ and $b_2$ represent $b(1)$ and $b(2)$, respectively. As a result, $\operatorname{dim}(\mathcal{S}(r))$ is equal to the number of linearly independent vectors in (\ref{Eq:four dependent vectors}). 

Then, we claim that the sequence of vectors $r, r^{\prime}, \mathbf{1}^{\prime}, \mathbf{1}^{\prime \prime}$ is linearly independent. To see this, we only need to show that the first 4 rows of column vectors $r, r^{\prime}, \mathbf{1}^{\prime}, \mathbf{1}^{\prime \prime}$ are linearly independent. Taking the first 4 rows of vectors $r, r^{\prime}, \mathbf{1}^{\prime}, \mathbf{1}^{\prime \prime}$, after some manipulation, we obtain
\small{
\begin{equation}
\label{Eq:first 4 dependent vectors}
\left[\begin{array}{@{}cccc@{}}
1 & 0 & 1 & 0 \\
0 & 1 & 0 & 1 \\
r_{11} & r_{12} & r_{21} & r_{22} \\
-r_{12} & r_{11} & -r_{22} & r_{21}
\end{array}\right]
\end{equation} }

Using \textit{Gaussian elimination}, after some manipulation, (\ref{Eq:first 4 dependent vectors}) is reduced to the following form
\small{
\begin{equation}
\label{Eq:U}
U=\left[\begin{array}{@{}ll:ll@{}}
1 & 0 & 1 & 0 \\
0 & 1 & 0 & 1 \\
\hdashline 0 & 0 & r_{21}-r_{11} & r_{22}-r_{12} \\
0 & 0 & r_{12}-r_{22} & r_{21}-r_{11}
\end{array}\right]
\end{equation} }

Note that 
\small {
\begin{equation}
\operatorname{det}\left[\begin{array}{@{}ll@{}}
r_{21}-r_{11} & r_{22}-r_{12} \\
r_{12}-r_{22} & r_{21}-r_{11}
\end{array}\right]=\left(r_{21}-r_{11}\right)^2+\left(r_{22}-r_{12}\right)^2
\end{equation} }
It follows from $r_i \neq r_j$ that $\operatorname{det}\left[\begin{array}{@{}ll@{}}r_{21}-r_{11} & r_{22}-r_{12} \\ r_{12}-r_{22} & r_{21}-r_{11}\end{array}\right] \neq 0$, which implies that matrix $U$ has full rank, and then directly leads to the assertion that the sequence of vectors $r, r^{\prime}, \mathbf{1}^{\prime}, \mathbf{1}^{\prime \prime}$ is linearly independent. Hence, our claim is established. That is, the dimension of $\mathcal{S}(r)$ equals 4 and $\mathcal{S}(r)=\operatorname{span}\left\{r, r^{\prime}, \mathbf{1}^{\prime}, \mathbf{1}^{\prime \prime}\right\}=\operatorname{span}\left\{r, r^{\prime}, \mathbf{1}_n \otimes I_2\right\}$.
\end{IEEEproof}

Motivated by Lemma \ref{lem:Dimension of similar image}, we make the following assumption on the nominal formation.
\begin{assumption}
\label{ass:no two nodes are collocated}
 The $r_i \neq r_j$ for $i, j \in \mathcal{V}$ and $n \geq 2$ are satisfied, i.e., no two nodes are collocated in the nominal formation $(\mathcal{G}, r)$.
\end{assumption}

\subsection{Similar localizability and leader selection}
Next, we discuss the selection of the number of leaders, i.e., how many leaders are needed for similar formation problem. First, we define a notion called similar localizability.

\begin{definition}[Similar localizability]
\label{def:Similar localizability}
The nominal formation $(\mathcal{G}, r)$ is similar localizable by the leaders if for any $p=\left[p_l^T, p_f^T\right]^T \in \mathcal{S}(r), p_f$ can be uniquely determined by $p_l$.  
\end{definition}

Based on Definition \ref{def:Similar localizability}, we give a necessary and sufficient condition of similar localizability.

\begin{theorem}[Leader selection for similar localizability]
\label{theo:Leader selection for similar localizability}
Under Assumption \ref{ass:no two nodes are collocated}, the nominal formation $(\mathcal{G}, r)$ is similar localizable if and only if $n_l \geq 2$ and $p_l \in \mathcal{S}\left(r_l\right)$. 
\end{theorem}
\begin{IEEEproof}[Proof]
For any $p \in \mathcal{S}(r)$, there exists $\alpha, \mathcal{R}, b$ such that
\begin{equation}
\label{Eq:Leader selection on scalar}
\left\{\begin{array}{@{}l}
p_1=\alpha \mathcal{R} \cdot r_1+b \\
\quad \vdots \\
p_n=\alpha \mathcal{R} \cdot r_n+b
\end{array}\right.
\end{equation}

With the notation of $\mathcal{R}$, after some manipulation, (\ref{Eq:Leader selection on scalar}) can be rewritten as the matrix form
\begin{equation}
\label{Eq:Leader selection on vector}
\underbrace{\left[\begin{array}{@{}c@{}}
p_1 \\
\vdots \\
p_n
\end{array}\right]}_p=\underbrace{\left[\begin{array}{@{}l:l@{}}
S_{r_1} & I_2 \\
\vdots & \vdots \\
S_{r_n} & I_2
\end{array}\right]}_{\bar{p}(r)}
\underbrace{\left[\begin{array}{@{}l@{}}
\alpha \mathcal{R}_{11} \\
\alpha \mathcal{R}_{21} \\ \hdashline[3pt/3pt]
\quad b
\end{array}\right]}_z
\end{equation}
where $S_{r_i}=\left[\begin{array}{@{}cc@{}}r_{i 1} & -r_{i 2} \\ r_{i 2} & r_{i 1}\end{array}\right], i=1, \ldots, n$. To further illustrate, (\ref{Eq:Leader selection on vector}) can be partitioned to be
\begin{equation}
\label{Eq:leaders part for localizability}
p_l =\bar{p}\left(r_l\right) z 
\end{equation}
\begin{equation}
\label{Eq:followers part for localizability}
p_f =\bar{p}\left(r_f\right) z
\end{equation}

(Sufficiency) Under $n_l \geq 2$ and Assumption \ref{ass:no two nodes are collocated}, it follows from Lemma \ref{lem:Dimension of similar image} that $\bar{p}\left(r_l\right)$ has full column rank, i.e., $\operatorname{rank}\left(\bar{p}\left(r_l\right)\right)=4$. Hence, (\ref{Eq:leaders part for localizability}) has $1$ or $0$ solution [\citenum{strang2022introduction}, Chapter 3]. If (\ref{Eq:leaders part for localizability}) is solvable, then $z$ can be uniquely determined as
\begin{equation*}
z=\left[\left(\bar{p}^T\left(r_l\right) \bar{p}\left(r_l\right)\right)^{-1} \bar{p}^T\left(r_l\right)\right] p_l
\end{equation*}
where it follows from the fact that $\operatorname{rank}\left(\bar{p}^T\left(r_l\right) \bar{p}\left(r_l\right)\right)=\operatorname{rank}\left(\bar{p}\left(r_l\right)\right)=4$ that matrix $\bar{p}^T\left(r_l\right) \bar{p}\left(r_l\right)$ is nonsingular. Also note that there is a unique corresponding relationship between $z$ and $\alpha, \mathcal{R}, b$ and thus it follows that the parameters $\alpha, \mathcal{R}, b$ can be uniquely determined. Consequently, $p_f$ can be uniquely determined using (\ref{Eq:Leader selection on scalar}) or (\ref{Eq:followers part for localizability}), which implies that the nominal formation is similar localizable. (Necessity) Now suppose by contradiction that $n_l=1$, then (\ref{Eq:leaders part for localizability}) will have infinitely many solutions for $z$, and hence the nominal formation is not similar localizable, a contradiction. Next, suppose that $p_l \notin \mathcal{S}\left(r_l\right)$, i.e., $p_l \notin \operatorname{span}\left\{r_l, r_l^{\prime}, \mathbf{1}_{n_l} \otimes I_2\right\}$. From Assumption \ref{ass:no two nodes are collocated} and the fact of $n_l \geq 2$, we know that Range $\left(\bar{p}\left(r_l\right)\right)=\operatorname{span}\left\{r_l, r_l^{\prime}, \mathbf{1}_{n_l} \otimes I_2\right\}$, which indicates that (\ref{Eq:leaders part for localizability}) has no solution for $z$ when $p_l \notin \operatorname{span}\left\{r_l, r_l^{\prime}, \mathbf{1}_{n_l} \otimes I_2\right\}$. Hence, $p_f$ cannot be uniquely determined, and thus the nominal formation is not similar localizable, a contradiction.
\end{IEEEproof}

Theorem \ref{theo:Leader selection for similar localizability} suggests that the number of leaders has at least two and the movement for leaders only be a similar movement, i.e., $p_l \in \operatorname{span}\left\{r_l, r_l^{\prime}, \mathbf{1}_{n_l} \otimes I_2\right\}$. When there are exactly two leaders, (\ref{Eq:leaders part for localizability}) is always solvable for any (not collocated) leader position. When there are more than two leaders, (\ref{Eq:leaders part for localizability}) is solvable if only if $p_l \in \operatorname{span}\left\{r_l, r_l^{\prime}, \mathbf{1}_{n_l} \otimes I_2\right\}$. That is, for the tall matrix $\bar{p}\left(r_l\right)$, (\ref{Eq:leaders part for localizability}) has a unique solution for $z$ when $p_l \in \operatorname{span}\left\{r_l, r_l^{\prime}, \mathbf{1}_{n_l} \otimes I_2\right\}$ is satisfied. Hence, two leaders are enough to obtain unique $z$, and more leaders are redundant, which motivates the following assumption.
\begin{assumption}
\label{ass:the number of leaders}
The number of leaders of the nominal formation $(\mathcal{G}, r)$ is two, i.e., $n_l=2$.
\end{assumption}

The following theorem provides a necessary and sufficient algebraic condition for similar localizability.
\begin{theorem}[Algebraic condition for similar localizability]
\label{theorem:Algebraic condition for similar localizability}
Under Assumption \ref{ass:no two nodes are collocated} and \ref{ass:the number of leaders}, the nominal formation $(\mathcal{G}, r)$ is similar localizable if and only if $\mathcal{L}_{f f}$ is nonsingular. When $\mathcal{L}_{f f}$ is nonsingular, for any $p=\left[p_l^T, p_f^T\right]^T \in \mathcal{S}(r), p_f$ can be uniquely determined by $p_f=-\mathcal{L}_{f f}^{-1} \mathcal{L}_{f l} p_l$.
\end{theorem}
\begin{IEEEproof}[Proof] 
(Sufficiency) First, it follows from (\ref{Eq:matrix-weighted Laplacian}) that $\mathcal{L} \cdot\left(\mathbf{1}_n \otimes I_2\right)=0$, which leads to the assertion of $\mathcal{L} \cdot\left(\mathbf{1}_n \otimes b\right)=0$. Then, the fact that $\mathcal{B}_{i j} \cdot \alpha \mathcal{R}=\alpha \mathcal{R} \cdot \mathcal{B}_{i j}$ show that $\mathcal{L} \cdot\left(I_n \otimes \alpha \mathcal{R}\right) r=\left(I_n \otimes \alpha \mathcal{R}\right) \cdot \mathcal{L} r=0$ by noting that (\ref{Eq:linear constraint a compact form}). Combining the above arguments, we conclude that $\mathcal{L} \cdot\left[\left(I_n \otimes \alpha \mathcal{R}\right) r+\mathbf{1}_n \otimes b\right]=0$. Since $\mathcal{L}_{f f}$ is nonsingular, note from (\ref{Eq:graph Laplacian}) and Assumption \ref{ass:the number of leaders} that $\operatorname{dim}(\operatorname{Null}(\mathcal{L}))=4$, and it thus follows from the fact that $\mathcal{L} \cdot\left[\left(I_n \otimes \alpha \mathcal{R}\right) r+\mathbf{1}_n \otimes b\right]=0$ and Lemma \ref{lem:Dimension of similar image} that $\mathcal{S}(r)=\operatorname{Null}(\mathcal{L})$ under Assumptions \ref{ass:no two nodes are collocated} and \ref{ass:the number of leaders}. Hence, we can infer that $\mathcal{L} p=0$ for any $p \in \mathcal{S}(r)$, which implies $\mathcal{L}_{f f} p_f+\mathcal{L}_{f l} p_l=0$. Since $\mathcal{L}_{f f}$ is nonsingular, $p_f$ can be uniquely determined by $p_f=-\mathcal{L}_{f f}^{-1} \mathcal{L}_{f l} p_l$, i.e., the nominal formation is similar localizable. (Necessity) To see this, suppose by contradiction that $\mathcal{L}_{f f}$ is singular. From Definition \ref{def:Similar localizability}, one has $\mathcal{L}_{f f} p_f=-\mathcal{L}_{f l} p_l$. Obviously, $p_f$ has infinite solutions for given $p_l$, a contradiction.
\end{IEEEproof}

\section{Similar formation framework construction}
For any directed acyclic graphs, we present a sufficient condition for similar localizability. This is our main result. 
\begin{theorem}            
\label{theorem:similar localizable for directed acyclic graph}
For a directed acyclic graph, if Assumptions \ref{ass:neighbors number}, \ref{ass:no two nodes are collocated} and \ref{ass:the number of leaders} are satisfied, then the given nominal formation $(\mathcal{G}, r)$ is similar localizable.
\end{theorem}
\begin{IEEEproof}[Proof] 
According to Lemma \ref{lem:reorder topology}, after renumbering the nodes $3 \sim n$ via a corresponding permutation matrix $P$, one has
\small{
\begin{equation*}
\mathcal{L}^{\prime}=\left[\begin{array}{@{}c:c@{}}
\mathbf{0} & \mathbf{0} \\
\hdashline \mathcal{L}_{f l}^{\prime} & \mathcal{L}_{f f}^{\prime}
\end{array}\right]=\left[\begin{array}{@{}c:c@{}}
I_4 & \mathbf{0} \\
\hdashline \mathbf{0} & \bar{P}
\end{array}\right]\left[\begin{array}{@{}c:c@{}}
\mathbf{0} & \mathbf{0} \\
\hdashline \mathcal{L}_{f l} & \mathcal{L}_{f f}
\end{array}\right]\left[\begin{array}{@{}c:c@{}}
I_4 & \mathbf{0} \\
\hdashline \mathbf{0} & \bar{P}^T
\end{array}\right]
\end{equation*}   }
where $\bar{P}=P \otimes I_2$ is also a permutation matrix. Without loss of generality, we assume that the Laplacain $\mathcal{L}^{\prime}$ (after renumbering) is of the following block lower triangular structure and all entries of $\mathcal{L}^{\prime}$ are configured as follows (the other case is similar for the next proof)
{\small
\begin{equation*}
\begin{aligned}
\mathcal{L}^{\prime}& =\left[\begin{array}{@{}c:c@{}}
\mathbf{0}_{4 \times 4} & \mathbf{0}_{4 \times 2 n_f} \\
\hdashline \mathcal{L}_{f l}^{\prime} & \mathcal{L}_{f f}^{\prime}
\end{array}\right]    \\
 &=\left[\begin{array}{@{}cc:cccc@{}}
\mathbf{0}_{2 \times 2} & \mathbf{0}_{2 \times 2} & \mathbf{0}_{2 \times 2} & \mathbf{0}_{2 \times 2} & \ldots & \mathbf{0}_{2 \times 2} \\
\mathbf{0}_{2 \times 2} & \mathbf{0}_{2 \times 2} & \mathbf{0}_{2 \times 2} & \mathbf{0}_{2 \times 2} & \ldots & \mathbf{0}_{2 \times 2} \\
\hdashline \mathcal{B}_{31}^{\prime} & \mathcal{B}_{32}^{\prime} & \mathcal{B}_{33}^{\prime} & \mathbf{0}_{2 \times 2} & \ldots & \mathbf{0}_{2 \times 2} \\
\mathcal{B}_{41}^{\prime} & \mathbf{0}_{2 \times 2} & \mathcal{B}_{43}^{\prime} & \mathcal{B}_{44}^{\prime} & \ldots & \mathbf{0}_{2 \times 2} \\
\vdots & \vdots & \vdots & \vdots & \ddots & \vdots \\
\mathcal{B}_{n 1}^{\prime} & \mathbf{0}_{2 \times 2} & \mathbf{0}_{2 \times 2} & \mathcal{B}_{n 4}^{\prime} & \cdots & \mathcal{B}_{n n}^{\prime}
\end{array}\right]
\end{aligned}
\end{equation*}
}
where $\mathcal{B}_{i j}^{\prime}$ is corresponding matrix-weighted value attributed on edge $(i, j)$ after renumbering nodes. Also reorder the components $r_i$ of the nominal configuration $r$ according to the above renumbering, and denote the result by $r^{\prime}$. Substituting the renumbered $\mathcal{L}^{\prime}$ and $r^{\prime}$ into equations $\mathcal{L} \cdot r=0$ and $\mathcal{L} \cdot\left(\mathbf{1}_n \otimes I_2\right)=0$ yields
\begin{equation}
\label{Eq:the renumbered Laplacian}
\left\{\begin{array}{@{}l@{}}
\mathcal{L}^{\prime} \cdot r^{\prime}=0 \\
\mathcal{L}^{\prime} \cdot\left(\mathbf{1}_n \otimes I_2\right)=0
\end{array}\right.
\end{equation}

For the 4th subblock row of $\mathcal{L}^{\prime}$ (other subblock rows are similar), it follows from (\ref{Eq:the renumbered Laplacian}) that
\begin{equation}
\label{Eq:linear equation for 4th}
\left\{\begin{array}{@{}l}
\mathcal{B}_{41}^{\prime}+\mathcal{B}_{43}^{\prime}+\mathcal{B}_{44}^{\prime}=0 \\
\mathcal{B}_{41}^{\prime} r_1^{\prime}+\mathcal{B}_{43}^{\prime} r_3^{\prime}+\mathcal{B}_{44}^{\prime} r_4^{\prime}=0
\end{array}\right.
\end{equation}

Note that the solutions to \small { $\mathcal{B}_{41}^{\prime}=\left[\begin{array}{@{}cc@{}}a_{41}^{\prime} & -b_{41}^{\prime} \\ b_{41}^{\prime} & a_{41}^{\prime}\end{array}\right], \mathcal{B}_{43}^{\prime}=\left[\begin{array}{@{}cc@{}}a_{43}^{\prime} & -b_{43}^{\prime} \\ b_{43}^{\prime} & a_{43}^{\prime}\end{array}\right]$ and $\mathcal{B}_{44}^{\prime}=\left[\begin{array}{@{}cc@{}}a_{44}^{\prime} & -b_{44}^{\prime} \\ b_{44}^{\prime} & a_{44}^{\prime}\end{array}\right]$ 
}
is equivalent to solving $a_{41}^{\prime}, b_{41}^{\prime}, a_{43}^{\prime}, b_{43}^{\prime}, a_{44}^{\prime}, b_{44}^{\prime}$. Hence, after some computation, it follows that (\ref{Eq:linear equation for 4th}) is equivalent to
\small{
\begin{equation}
\label{Eq:equivalent form}
\left[\begin{array}{@{}cccccc@{}}
1 & 0 & 1 & 0 & 1 & 0 \\
0 & 1 & 0 & 1 & 0 & 1 \\
r_{11}^{\prime} & -r_{12}^{\prime} & r_{31}^{\prime} & -r_{32}^{\prime} & r_{41}^{\prime} & -r_{42}^{\prime} \\
r_{12}^{\prime} & r_{11}^{\prime} & r_{32}^{\prime} & r_{31}^{\prime} & r_{42}^{\prime} & r_{41}^{\prime}
\end{array}\right] \cdot\left[\begin{array}{@{}c@{}}
a_{41}^{\prime} \\
b_{41}^{\prime} \\
a_{43}^{\prime} \\
b_{43}^{\prime} \\
a_{44}^{\prime} \\
b_{44}^{\prime}
\end{array}\right]=0
\end{equation} }

Then, perform the following elementary row transformation (\textit{Gaussian elimination})
\small {
\begin{equation*}
\begin{aligned}
& {\left[\begin{array}{@{}cccccc@{}}
1 & 0 & 1 & 0 & 1 & 0 \\
0 & 1 & 0 & 1 & 0 & 1 \\
r_{11}^{\prime} & -r_{12}^{\prime} & r_{31}^{\prime} & -r_{32}^{\prime} & r_{41}^{\prime} & -r_{42}^{\prime} \\
r_{12}^{\prime} & r_{11}^{\prime} & r_{32}^{\prime} & r_{31}^{\prime} & r_{42}^{\prime} & r_{41}^{\prime}
\end{array}\right] \xrightarrow{\substack{\text { Gaussian } \\
\text { elimination }}}} \\
& {\left[\begin{array}{@{}cccccc@{}}
1 & 0 & 1 & 0 & 1 & 0 \\
0 & 1 & 0 & 1 & 0 & 1 \\
0 & 0 & r_{31}^{\prime}-r_{11}^{\prime} & r_{12}^{\prime}-r_{32}^{\prime} & r_{41}^{\prime}-r_{11}^{\prime} & r_{12}^{\prime}-r_{42}^{\prime} \\
0 & 0 & r_{32}^{\prime}-r_{12}^{\prime} & r_{31}^{\prime}-r_{11}^{\prime} & r_{42}^{\prime}-r_{12}^{\prime} & r_{41}^{\prime}-r_{11}^{\prime}
\end{array}\right]}
\end{aligned}
\end{equation*}  }

Based on Assumption \ref{ass:no two nodes are collocated}, one has 
{\small
\begin{equation*}
\operatorname{det}\left[\begin{array}{@{}ll@{}}
r_{31}^{\prime}-r_{11}^{\prime} & r_{12}^{\prime}-r_{32}^{\prime} \\
r_{32}^{\prime}-r_{12}^{\prime} & r_{31}^{\prime}-r_{11}^{\prime}
\end{array}\right]=\left(r_{31}^{\prime}-r_{11}^{\prime}\right)^2+\left(r_{32}^{\prime}-r_{12}^{\prime}\right)^2 \neq 0
\end{equation*}
} which means all solutions to (\ref{Eq:equivalent form}) span a space that the dimension equals 2. To satisfy above equations, the entries $a_{41}^{\prime}, b_{41}^{\prime}, a_{43}^{\prime}, b_{43}^{\prime}, a_{44}^{\prime}, b_{44}^{\prime}$ are such that
\small {
\begin{equation}
\label{Eq:final form}
\left[\begin{array}{@{}l@{}}
a_{41}^{\prime} \\
b_{41}^{\prime} \\
a_{43}^{\prime} \\
b_{43}^{\prime} \\
a_{44}^{\prime} \\
b_{44}^{\prime}
\end{array}\right]=c_1^{\prime}\left[\begin{array}{@{}c@{}}
r_{31}^{\prime}-r_{41}^{\prime} \\
r_{32}^{\prime}-r_{42}^{\prime} \\
r_{41}^{\prime}-r_{11}^{\prime} \\
r_{42}^{\prime}-r_{12}^{\prime} \\
r_{11}^{\prime}-r_{31}^{\prime} \\
r_{12}^{\prime}-r_{32}^{\prime}
\end{array}\right]+c_2^{\prime}\left[\begin{array}{@{}l@{}}
r_{42}^{\prime}-r_{32}^{\prime} \\
r_{31}^{\prime}-r_{41}^{\prime} \\
r_{12}^{\prime}-r_{42}^{\prime} \\
r_{41}^{\prime}-r_{11}^{\prime} \\
r_{32}^{\prime}-r_{12}^{\prime} \\
r_{11}^{\prime}-r_{31}^{\prime}
\end{array}\right]
\end{equation}   }
where $c_1^{\prime}$ and $c_2^{\prime}$ are not all zero. It can be observed that $\left[a_{44}^{\prime}, b_{44}^{\prime}\right]^T \neq 0$ based on Assumption \ref{ass:no two nodes are collocated}. Using the fact that the matrix \small {
$\left[\begin{array}{@{}cc@{}}a_{44}^{\prime} & -b_{44}^{\prime} \\ b_{44}^{\prime} & a_{44}^{\prime}\end{array}\right]$  }
is singular if only if $\left[a_{44}^{\prime}, b_{44}^{\prime}\right]^T=0$, we infer that $\mathcal{B}_{44}^{\prime}$ is nonsingular from (\ref{Eq:final form}), which further concludes all $\mathcal{B}_{i i}^{\prime}(i=3, \ldots, n)$ are nonsingular using similar steps. As a result, one has $\mathcal{L}_{f f}^{\prime}$ is nonsingular and it thus follows from the fact that $\mathcal{L}_{f f}^{\prime}=\bar{P} \mathcal{L}_{f f} \bar{P}^T$ and $\bar{P}^{-1}=\bar{P}^T$ that $\mathcal{L}_{f f}$ is nonsingular. According to Definition \ref{def:Similar localizability}, the given nominal $(\mathcal{G}, r)$ is similar localizable.    
\end{IEEEproof}

Note that the eigenvalues of $\mathcal{L}_{f f}$ may not be located entirely in the right half-plane. Then, we make the following claim.

For a directed acyclic graph, the following transformation can make all eigenvalues of $\mathcal{L}_{f f}$ have positive real parts
\small {
\begin{equation}
\label{Eq:update Laplacain}
\mathcal{L}=\left[\begin{array}{@{}c:c@{}}
I_4 & \mathbf{0} \\
\hdashline \mathbf{0} & E
\end{array}\right] \cdot \mathcal{L}
\end{equation}  }
where $E=\operatorname{diag}\left(E_{33}, \ldots, E_{n n}\right), E_{i i}=\mathcal{B}^T{ }_{i i} / c_{i i}^2$ ($c_{ii}$ is defined in (\ref{Eq:matrix-weighted value})), $i=3, \ldots, n$. To see this, for a directed acyclic graph, since a block lower triangular matrix can be obtained from permutation similarity transformation $\bar{P} E \mathcal{L}_{f f} \bar{P}^T$, but their block diagonal elements set $\left\{E_{33} \mathcal{B}_{33}, \ldots, E_{n n} \mathcal{B}_{n n}\right\}$ remains unchanged (possible change in order) via permutation similarity transformation. Note that the fact that the eigenvalues of a block lower triangular matrix are the union of the eigenvalues of all its diagonal block matrices, so it follows that all eigenvalues of block lower triangular matrix $E \cdot \mathcal{L}_{f f}$ have positive real parts if only if the eigenvalues of all block diagonal elements of $\bar{P} E \mathcal{L}_{f f} \bar{P}^T$ have positive real parts. It is obvious that all eigenvalues of $\bar{P} E \mathcal{L}_{f f} \bar{P}^T$ equal 1. Hence, our claim is established.

\section{Similar formation control in 2-D plane}
Consider $n$ agents with single-integrator dynamics in 2-D plane given by
\begin{equation}
\label{Eq:single-integrator dynamics}
\dot{p}_i=u_i, \quad i=1, \ldots, n
\end{equation}
where $u_i(t) \in \mathbb{R}^2$ is the control input.

To achieve a similar formation, consider the following distributed control law
\begin{equation}
\label{Eq:distributed control law}
u_i=-\sum_{j \in \mathcal{N}_i} \mathcal{B}_{i j}\left(p_i-p_j\right)
\end{equation}

Using (\ref{Eq:matrix-weighted Laplacian}) and (\ref{Eq:distributed control law}), (\ref{Eq:single-integrator dynamics}) can be written in vector form as 
\begin{equation}
\label{Eq:vector form with single-integrator dynamics}
\dot{p}=-\mathcal{L} \cdot p
\end{equation}

Note that $-\mathcal{L}$ can be written in Jordan canonical form as
\small{
\begin{equation}
\label{Eq:Jordan canonical form}
\begin{aligned}
& -\mathcal{L}=M J M^{-1}= \\
& {\left[\mathbf{1}^{\prime}, \mathbf{1}^{\prime \prime}, r, r^{\prime}, y_3, \ldots, y_{2 n}\right]\left[\begin{array}{@{}c:c@{}}
\mathbf{0}_{4 \times 4} & \mathbf{0}_{4 \times(2 n-4)} \\
\hdashline \mathbf{0}_{(2 n-4) \times 4} & J^{\prime}
\end{array}\right]\left[\begin{array}{@{}c@{}}
z_1^T \\
\vdots \\
z_{2 n}^T
\end{array}\right]}
\end{aligned}
\end{equation}
} where $y_i \in \mathbb{R}^{2 n}(i=3, \ldots n)$ is the (generalized) right eigenvectors of $-\mathcal{L}, z_i \in \mathbb{R}^{2 n}(i=1, \ldots, n)$ is the (generalized) left eigenvectors of $-\mathcal{L}$. It follows from (\ref{Eq:update Laplacain}) that all the eigenvalues of $-\mathcal{L}_{ff}$ have positive real parts, which implies that all the eigenvalues of $J^{\prime}$ have negative real parts. Hence, we see that
\begin{equation}
\begin{aligned}
&e^{-\mathcal{L} \cdot t}  =e^{M J M^{-1} t}=M e^{J \cdot t} M^{-1} \\
& =M\left[\begin{array}{c:c}
I_4 & \mathbf{0}_{4 \times(2 n-4)} \\
\hdashline \mathbf{0}_{(2 n-4) \times 4} & e^{J^{\prime} \cdot t}
\end{array}\right] M^{-1} \\
& \rightarrow \mathbf{1}^{\prime} \cdot z_1^T+\mathbf{1}^{\prime \prime} \cdot z_2^T+r \cdot z_3^T+r^{\prime} \cdot z_4^T, \text { as } t \rightarrow \infty
\end{aligned}
\end{equation}

Thus, it follows that the solution of (\ref{Eq:vector form with single-integrator dynamics}) is
\begin{equation}
\begin{aligned}
 p(t) & =e^{-\mathcal{L} \cdot t} p(0) \\
& \rightarrow \mathbf{1}^{\prime} \cdot z_1^T p(0)+\mathbf{1}^{\prime \prime} \cdot z_2^T p(0)\\
& \quad +r \cdot z_3^T p(0)+r^{\prime} \cdot z_4^T p(0),\text { as } t \rightarrow \infty
\end{aligned}
\end{equation}

Let $\bar{r}:=\mathbf{1}^{\prime} \cdot z_1^T p(0)+\mathbf{1}^{\prime \prime} \cdot z_2^T p(0)+r \cdot z_3^T p(0)+r^{\prime} \cdot z_4^T p(0)$, it follows from Lemma \ref{lem:Dimension of similar image} that $\bar{r} \in \mathcal{S}(r)$ i.e.,
\begin{equation}
\lim _{x \rightarrow \infty} p(t) \in \mathcal{S}(r)
\end{equation}
which is equivalent to $\mathcal{L} \cdot p(\infty) \rightarrow 0 \quad$ by noting that $\mathcal{S}(r)=\operatorname{Null}(\mathcal{L})$, i.e., $\mathcal{L}_{f f} p_f(\infty)+\mathcal{L}_{f l} p_l(\infty) \rightarrow 0$. Furthermore, one has
\begin{equation}
\lim _{t \rightarrow \infty}\left(p_f(t)+\mathcal{L}_{f f}^{-1} \mathcal{L}_{f l} p_l(t)\right) \rightarrow 0
\end{equation}
Namely the control object (\ref{Eq:control objective 2}) is achieved.

\section{Simulation}
\begin{figure}[!t]
\centering
\includegraphics[width=1.8in]{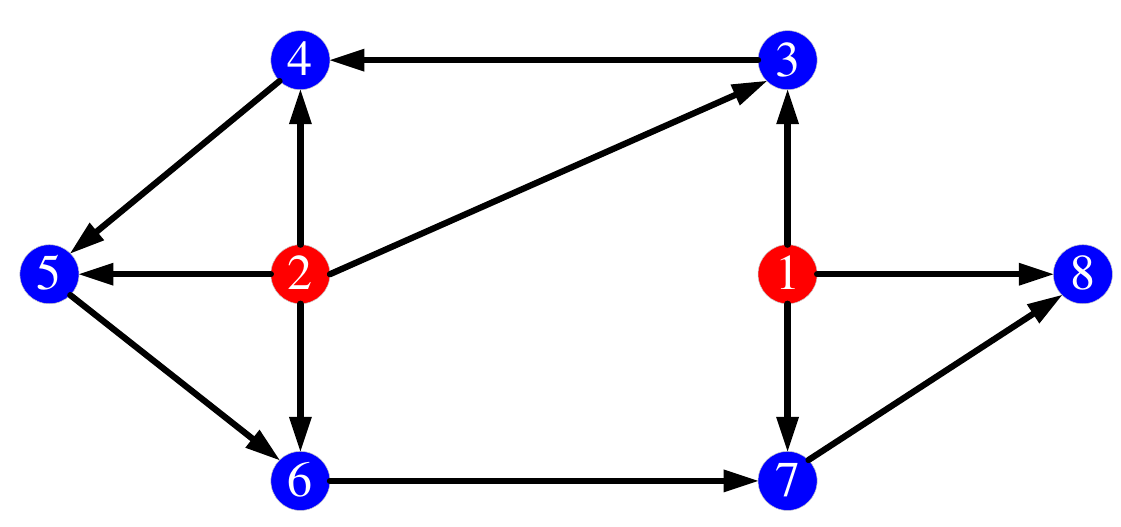}
\caption{The communication topological graph.}
\label{fig:topology}
\end{figure}

In the simulation, the proposed similar formation method is applied to the control of single integrator dynamics. Multi-agent systems consist of two leaders $\mathcal{V}_l=\{1,2\}$ and six followers $\mathcal{V}_f=\{3,4,5,6,7,8\}$. The communication topological graph is shown in Fig. \ref{fig:topology}.

The localizable nominal configuration is designed as (see Fig. \ref{fig:topology})
{\small
\begin{equation*}
\begin{aligned}
& r_1^T=[1,0]^T, r_2^T=[-1,0]^T, r_3^T=[1,1]^T,r_4^T=[-1,1]^T, \\
&  r_5^T=[-2,0]^T, r_6^T=[-1,-1]^T, r_7^T=[1,-1]^T, r_8^T=[2,0]^T
\end{aligned}
\end{equation*}
}

Based on (\ref{Eq:equilibrium condition}) and (\ref{Eq:update Laplacain}), the matrix-weighted value is selected as
{\small
\begin{equation*}
\begin{aligned}
& \mathcal{B}_{31}=\left[\begin{array}{@{}cc@{}}
-1 & 0.5 \\
-0.5 & -1
\end{array}\right], \mathcal{B}_{32}=\left[\begin{array}{@{}cc@{}}
0 & -0.5 \\
0.5 & 0
\end{array}\right], \mathcal{B}_{33}=\left[\begin{array}{@{}ll@{}}
1 & 0 \\
0 & 1
\end{array}\right] \\
& \mathcal{B}_{42}=\left[\begin{array}{@{}cc@{}}
-0.8 & -0.4 \\
0.4 & -0.8
\end{array}\right], \mathcal{B}_{43}=\left[\begin{array}{@{}cc@{}}
-0.2 & 0.4 \\
-0.4 & -0.2
\end{array}\right], \mathcal{B}_{44}=\left[\begin{array}{@{}ll@{}}
1 & 0 \\
0 & 1
\end{array}\right] \\
& \mathcal{B}_{52}=\left[\begin{array}{@{}cc@{}}
-1 & -1 \\
1 & -1
\end{array}\right],   \mathcal{B}_{54}=\left[\begin{array}{@{}cc@{}}
0 & 1 \\
-1 & 0
\end{array}\right], \mathcal{B}_{55}=\left[\begin{array}{@{}ll@{}}
1 & 0 \\
0 & 1
\end{array}\right] \\
& \mathcal{B}_{62}=\left[\begin{array}{@{}cc@{}}
-1 & -1 \\
1 & -1
\end{array}\right], \mathcal{B}_{65}=\left[\begin{array}{@{}cc@{}}
0 & 1 \\
-1 & 0
\end{array}\right], \mathcal{B}_{66}=\left[\begin{array}{@{}ll@{}}
1 & 0 \\
0 & 1
\end{array}\right] \\
& \mathcal{B}_{71}=\left[\begin{array}{@{}cc@{}}
-0.8 & -0.4 \\
0.4 & -0.8
\end{array}\right], \mathcal{B}_{76}=\left[\begin{array}{@{}cc@{}}
-0.2 & 0.4 \\
-0.4 & -0.2
\end{array}\right], \mathcal{B}_{77}=\left[\begin{array}{@{}ll@{}}
1 & 0 \\
0 & 1
\end{array}\right] \\
& \mathcal{B}_{81}=\left[\begin{array}{@{}cc@{}}
-1 & -1 \\
1 & -1
\end{array}\right],   \mathcal{B}_{87}=\left[\begin{array}{@{}cc@{}}
0 & 1 \\
-1 & 0
\end{array}\right], \mathcal{B}_{88}=\left[\begin{array}{@{}cc@{}}
1 & 0 \\
0 & 1
\end{array}\right]
\end{aligned}
\end{equation*}
}

\begin{figure}[H]
    \centering
    \subfloat[\tiny ]{%
        \includegraphics[width=0.5\linewidth, height=3.3cm]{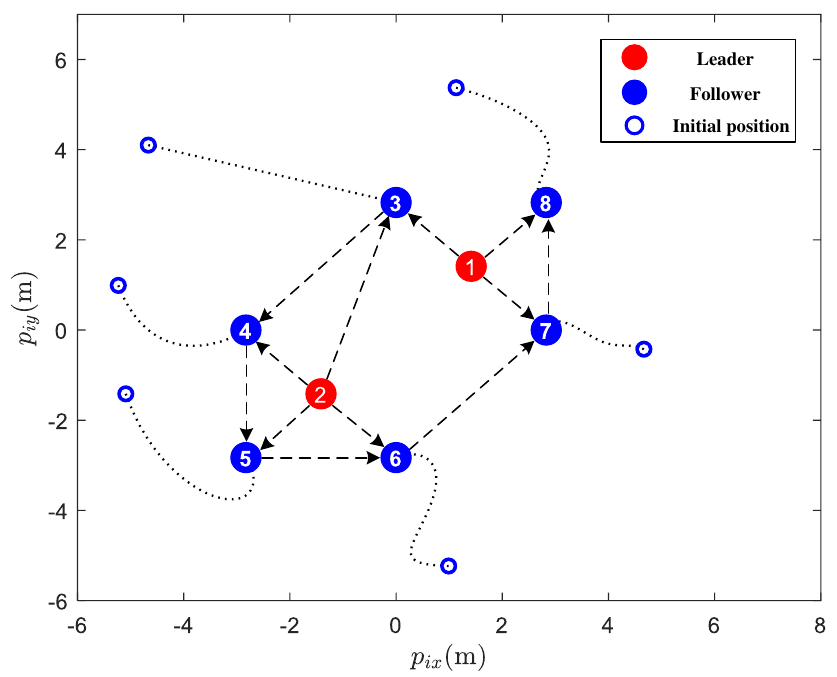}%
        \label{fig:1}%
    }\hfill
    \subfloat[\tiny]{%
        \includegraphics[width=0.5\linewidth, height=3.3cm]{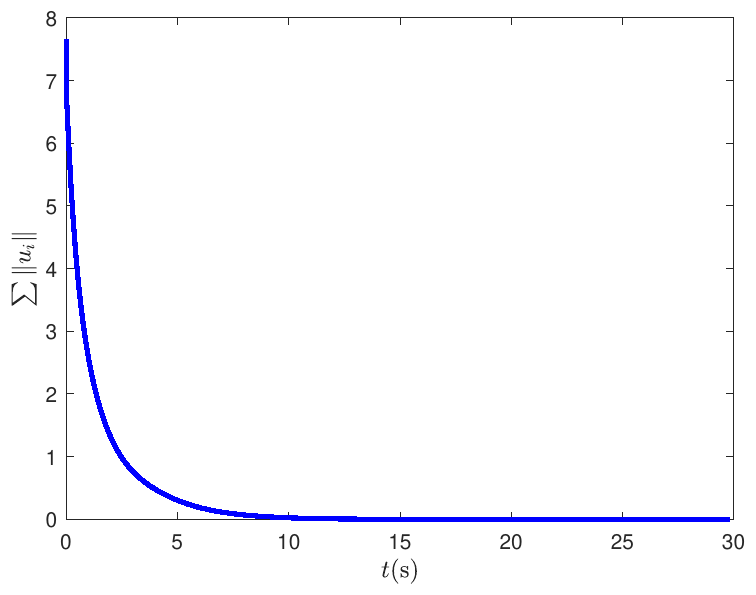}%
        \label{fig:2}%
    }
    \caption{(a) Multi-agent trajectory evolution. (b) System control input.}  
    \label{fig:simulition results}   
\end{figure}

Fig. \ref{fig:simulition results}(a) shows the trajectories of eight agents starting from random initial states, where a similar formation is successfully achieved. Observe that the two leaders 1 and 2 have stayed put as their initial and final positions coincide, that is, because they have no neighbors and thus have never updated their position. Fig. \ref{fig:simulition results}(b) shows that the system control input ultimately converges to zero, which means $\lim _{x \rightarrow \infty} p(t) \in \mathcal{S}(r)$.    

\section{Conclusion}
This brief solves the issue of similar formation control over directed acyclic graphs. The sufficient graphical condition for similar localizability is given, and the leader selection problem is addressed. However, there are several important topics that deserve further study in the future. For example, the results presented in this brief may be generalized by considering more complicated agent dynamics and motion constraints. 

\bibliographystyle{IEEEtran}
\bibliography{main}

\end{document}